\RequirePackage{fix-cm}
\documentclass[twocolumn,epjc3]{svjour3}  
\smartqed  % flush right qed marks, e.g. at end of proof
\RequirePackage[dvipdfmx]{graphicx}
\RequirePackage{latexsym}
\RequirePackage[numbers,sort&compress]{natbib}
\RequirePackage[colorlinks,citecolor=blue,urlcolor=blue,linkcolor=blue]{hyperref}
\usepackage{array}
\usepackage{braket}
\usepackage{bm}
\usepackage{subfigure}
\usepackage{verbatim}
\usepackage{wrapfig}
\usepackage{ascmac}
\usepackage{makeidx}
\usepackage{ulem}
\usepackage[version=3]{mhchem}
\usepackage{here}
\usepackage{comment}
\newcommand{\JSNS}{JSNS$^2$\,}

\newcommand{\nuebar}{$\bar{\nu}_{e}$~}

\newcommand{\Sterile}{$\bar{\nu}_{\mu} \to \bar{\nu}_{e}$~}

\newcommand{\IBD}{$\bar{\nu}_{e} + p \to e^{+} + n$}

\newcommand{\micro}{$\mu \mathrm{s}$\,}

\newcommand{\eVsq}{$\mathrm{eV}^2$}

\journalname{Eur. Phys. J. C}
\begin{document}

\title{
Characterization of the correlated background for a sterile neutrino search using the first dataset of the \JSNS experiment
}

\author{
Y.~Hino\thanksref{e1, addr4} \and
S.~Ajimura\thanksref{addr1} \and
M.~K.~Cheoun\thanksref{addr2} \and
J.~H.~Choi\thanksref{addr3} \and
T.~Dodo\thanksref{addr4, addr23} \and
H.~Furuta\thanksref{addr4} \and
J.~Goh\thanksref{addr22} \and
K.~Haga\thanksref{addr5} \and
M.~Harada\thanksref{addr5} \and
S.~Hasegawa\thanksref{addr23, addr5} \and
T.~Hiraiwa\thanksref{addr1} \and
W.~Hwang\thanksref{addr22} \and
H.~I.~Jang\thanksref{addr9} \and
J.~S.~Jang\thanksref{addr8} \and
H.~Jeon\thanksref{addr19} \and
S.~Jeon\thanksref{addr19} \and
K.~K.~Joo\thanksref{addr10} \and
J.~R.~Jordan\thanksref{addr6} \and
D.~E.~Jung\thanksref{addr19} \and
S.~K.~Kang\thanksref{addr11} \and
Y.~Kasugai\thanksref{addr5} \and
T.~Kawasaki\thanksref{addr7} \and
E.~J.~Kim\thanksref{addr12} \and
J.~Y.~Kim\thanksref{addr10} \and
S.~B.~Kim\thanksref{addr19} \and
W.~Kim\thanksref{addr14} \and
H.~Kinoshita\thanksref{addr5} \and
T.~Konno\thanksref{addr7} \and
C.~Y.~Lee\thanksref{addr2} \and %C.~Y.~Cheoun\thanksref{addr2} \and
D.~H.~Lee\thanksref{addr15} \and
S.~Lee\thanksref{addr22} \and
I.~T.~Lim\thanksref{addr10} \and
C.~Little\thanksref{addr6} \and
E.~Marzec\thanksref{addr6} \and
T.~Maruyama\thanksref{e2, addr15} \and
S.~Masuda\thanksref{addr5} \and
S.~Meigo\thanksref{addr5} \and
S.~Monjushiro\thanksref{addr15} \and
D.~H.~Moon\thanksref{addr10} \and
T.~Nakano\thanksref{addr1} \and
M.~Niiyama\thanksref{addr16} \and
K.~Nishikawa\thanksref{addr15} \and
M.~Nomachi\thanksref{addr1} \and
M.~Y.~Pac\thanksref{addr3} \and
J.~S.~Park\thanksref{addr15} \and
R.~G.~Park\thanksref{addr10} \and
S.~J.~M.~Peeters\thanksref{addr17} \and
H.~Ray\thanksref{addr18} \and
C.~Rott\thanksref{addr19, addr24} \and
K.~Sakai\thanksref{addr5} \and
S.~Sakamoto\thanksref{addr5} \and
T.~Shima\thanksref{addr1} \and
C.~D.~Shin\thanksref{addr10} \and
J.~Spitz\thanksref{addr6} \and
F.~Suekane\thanksref{addr4} \and
Y.~Sugaya\thanksref{addr1} \and
K.~Suzuya\thanksref{addr5} \and
M.~Taira\thanksref{addr15} \and
R.~Ujiie\thanksref{addr4} \and
Y.~Yamaguchi\thanksref{addr5} \and
M.~Yeh\thanksref{addr21} \and
I.~S.~Yeo\thanksref{addr3} \and %affil modified
C.~Yoo\thanksref{addr22} \and %added due to missing
I.~Yu\thanksref{addr19}
}

\thankstext{e1}{e-mail: hino@awa.tohoku.ac.jp}
\thankstext{e2}{e-mail: Takasumi.Maruyama@kek.jp}


\institute{%
Research Center for Neutrino Science, Tohoku University, Sendai, Miyagi, JAPAN \label{addr4} \and
Research Center for Nuclear Physics, Osaka University, Osaka, JAPAN \label{addr1} \and
Department of Physics, Soongsil University, Seoul 06978, KOREA \label{addr2} \and
Laboratory for High Energy Physics, Dongshin University, Chonnam 58245, KOREA \label{addr3} \and
Advanced Science Research Center, JAEA, Ibaraki JAPAN \label{addr23} \and
Department of Physics, Kyung Hee University, Seoul 02447, KOREA \label{addr22} \and
J-PARC Center, JAEA, Tokai, Ibaraki JAPAN \label{addr5} \and
Department of Fire Safety, Seoyeong University, Gwangju 61268, KOREA \label{addr9} \and
Gwangju Institute of Science and Technology, Gwangju, 61005, KOREA \label{addr8} \and
Department of Physics, Sungkyunkwan University, Gyeong Gi-do, KOREA \label{addr19} \and
Department of Physics, Chonnam National University, Gwangju, 61186, KOREA \label{addr10} \and
University of Michigan, Ann Arbor, MI, 48109, USA \label{addr6} \and
School of Liberal Arts, Seoul National University of Science and Technology, Seoul, 139-743, KOREA \label{addr11} \and
Department of Physics, Kitasato University, Sagamihara 252-0373, Kanagawa, JAPAN \label{addr7} \and
Division of Science Education, Jeonbuk National University, Jeonju, 54896, KOREA \label{addr12} \and
Department of Physics, Kyungpook National University, Daegu 41566, KOREA \label{addr14} \and
High Energy Accelerator Research Organization (KEK), Tsukuba, Ibaraki, JAPAN \label{addr15} \and
Department of Physics, Kyoto Sangyo University, Kyoto, JAPAN \label{addr16} \and
Department of Physics and Astronomy, University of Sussex, Brighton, UK \label{addr17} \and
University of Florida, Gainesville, FL, 32611, USA \label{addr18} \and
Department of Physics, The University of Utah, UT, 84112, USA \label{addr24} \and
University of Alabama, Tuscaloosa, AL, 35487, USA \label{addr20} \and % for Ion
Brookhaven National Laboratory, Upton, NY, 11973-5000, USA \label{addr21}% \and
%\emph{Present Address:} if needed\label{addr3}
}

\date{Received: date / Accepted: date}

\maketitle

\begin{abstract}
%Insert your abstract here. Include keywords, PACS and mathematical
%subject classification numbers as needed.
\JSNS (J-PARC Sterile Neutrino Search at J-PARC Spallation Neutron Source) is an experiment that is searching for sterile neutrinos via the observation of \Sterile appearance oscillations using muon decay-at-rest neutrinos.
Before dedicated data taking in the first-half of 2021, we performed a commissioning run for 10 days in June 2020.
Using the data obtained in this commissioning run, in this paper, 
we present an estimate of the correlated background which imitates the \nuebar signal in a sterile neutrino search.
In addition, 
in order to demonstrate future prospects of the \JSNS experiment, possible pulse shape discrimination improvements towards reducing cosmic ray induced fast neutron background are described.
\end{abstract}

\section{Introduction} %Your text comes here. Separate text sections with
\label{intro}
The existence of sterile neutrinos has been an important issue in the field of neutrino physics for over 20 years. The experimental results from~\cite{cite:LSND, cite:GALLEX, cite:SAGE, cite:RA, cite:MiniBooNE2013, cite:MiniBooNE2018} could be interpreted as indications of the existence of sterile neutrinos with mass-squared differences of around 1~\eVsq. %~with the three weakly interacting neutrinos

The \JSNS experiment, proposed in 2013~\cite{cite:proposal}, is designed to search for neutrino oscillations caused by such a sterile neutrino at the Material and Life science experimental Facility (MLF) in J-PARC.
The facility provides an intense and high-quality neutrino beam with $1.8 \times 10^{14}\,\nu$/year/cm$^2$ from muon decay-at-rest ($\mu$DAR)
produced using a 1~MW proton beam with a 25~Hz repetition rate.
The neutrino source is produced by impinging 3~GeV protons from a rapid cycling synchrotron on a mercury target in the MLF.
The experiment uses a Gadolinium (Gd) loaded liquid scintillator (Gd-LS) detector with 0.1~w\% Gd concentration placed at 24~m from the target.

The \JSNS experiment aims to directy test the LSND observation~\cite{cite:LSND} with improvements of the experimental technique.
Observing \Sterile oscillation using a $\mu$DAR neutrino source via inverse beta decay (IBD) reaction, \IBD, is the same experimental principle used by the LSND experiment~\cite{cite:LSND}.
On the other hand, there are several improvements offered by the \JSNS experiment.
In order to identify IBD events, a delayed coincidence between the positron signal (prompt signal: up to 53~MeV) and neutron capture signal is used for selection. Gd is used to identify neutron captures. After capturing thermal neutrons, Gd generates gamma-rays with higher energies and shorter capture times (8~MeV, $\sim30~\mu$s) than hydrogen (2.2~MeV, $\sim 200~\mu$s).
Therefore, backgrounds accidentally coincident in the delayed signal region can be reduced by $\sim 6$ times compared to the hydrogen capture used in the LSND experiment, due to the shorter capture time.
In addition, the short-pulsed beam, 100 ns width double pulses in 600 ns interval in each spill, enables us to set a timing window for the IBD prompt signal as 1.5 to 10~\micro~from the proton beam collision so that the neutrinos from pion and kaon decay and fast neutrons generated at the target can be rejected efficiently.
However, the efficiency for the $\mu$DAR neutrinos can be kept at 62\% because of the muon lifetime ($2.2~\mathrm{\mu s}$).
The cosmogenic background is also reduced by a factor of $10^{-4}$.

There are a number of correlated backgrounds to the IBD signal, characterized by a time-coincident prompt and delayed signal.
The most concerning one is a cosmic-induced neutron whose prompt signal is made by a recoil proton and a delayed signal from the neutron capture on Gd after thermalization. Approximately 99\% of this neutron background can be rejected by pulse shape discrimination (PSD) using the waveform shape difference between the prompt signal of IBD and neutron events.
Detailed discussion about the signal detection principle and the background rejection technique can be found in~\cite{cite:proposal, cite:TDR}.

\section{Commissioning Run}
\label{com_run}
The \JSNS experiment commenced data taking with a single detector in June 2020. 
The data taking period was from June 5th to 15th - approximately 10 days.
The proton beam power was 600~kW during this period. There was a one-day beam-off period due to facility maintenance, and thus we obtained beam-off data at that time.
The integrated number of proton-on-target (POT) collected was $8.9 \times 10^{20}$, corresponding to less than 1.0\% of the required POT of the \JSNS experiment. The expected number of IBD signal events from this run is estimated to be much less than 1~\cite{cite:TDR}. Therefore, all observed events are likely background.

\subsection{Experimental setup}
\label{setup}
The \JSNS detector is a cylindrical liquid scintillator detector with 4.6~m diameter and 3.5~m height placed at a distance of 24 m from the mercury target of the MLF. It consists of 17 tonnes of Gd-LS contained in an acrylic vessel, and 33 tonnes unloaded liquid scintillator (LS) in a layer between the acrylic vessel and a stainless steel tank.
The LS and the Gd-LS consist of LAB (linear alkyl benzene) as the base solvent, 3 g/L PPO (2,5-diphenyloxazole) as the fluor, and 15 mg/L bis-MSB (1,4-bis(2-methylstyryl) benzene) as the wavelength shifter.
The LS volume is separated into two independent layers by an optical separator that forms two detector volumes in the one detector.
The region inside the optical separator, called the ``inner detector'', consists of the entire volume of the Gd-LS and $\sim$25 cm thick LS layer.
Scintillation light from the inner detector is observed by 96 Hamamatsu R7081 photomultiplier tubes (PMTs) each with a 10-inch diameter.
The outer layer, called the ``veto layer'', is used to detect cosmic-ray induced particles coming into the detector.
A total of 24 of 10-inch PMTs are set in the veto layer whose inner surfaces are fully covered with reflection sheets in order to improve the collection efficiency of the scintillation light.

\subsection{Data acquisition system and triggers}
\label{daq}
PMT signal waveforms from both the inner detector and the veto layer are digitized and recorded at a 500~MHz sampling rate by 8-bit flash analog-to-digital converters (FADCs).
As a trigger, we utilize a 25 Hz periodic signal from the kicker magnet which directs the proton beam towards the MLF target, called the ``kicker trigger''.
The width of the acquired waveform in this trigger scheme is set to 25~\micro, which thoroughly covers the prompt signal timing window of the IBD events.
We mainly used this trigger and data acquisition to obtain the beam-on data for a background estimation within the sterile neutrino search.

The other trigger, called the ``self trigger'', uses an analog sum of the PMT signals from the inner detector. The trigger threshold is set to 80~mV equivalent to 100 p.e. and its efficiency reaches approximately 99\% above 7~MeV. The FADCs record 496~ns wide waveforms during the data taking using the self trigger. We utilized the self trigger during the beam-off period, in order to obtain cosmogenic events, and when taking calibration data using a 252-Californium ($^{252}$Cf) neutron source.
A detailed description of the detector and the triggers are given in~\cite{cite:detector, cite:daq}, respectively.

\section{Event selection and results}
\label{analysis}
The data set used for the background estimation was obtained using the kicker trigger.
The total number of beam spills was 8092503, equivalent to 3.5~days of data taking.
Since the obtained waveforms, given the 25~\micro width, contain multiple events,
we therefore used an event definition based on the number of hit PMTs in order to extract each event.
We constructed a hit time series at each trigger by accumulating hit information along the FADC window with 60~ns coincidence width over all the PMTs.
The coincidence width was determined by considering a typical PMT pulse shape and a safety factor from timing calibration.
%The safety factor from 8 ns resolution among PMT channels exists at the sequence of the event building.
The event discrimination threshold is set to 10 hits and 50 p.e., which corresponds to an energy well below 1~MeV.

The event vertex position and energy reconstruction is performed simultaneously based on a maximum-likelihood algorithm using the charge response of each PMT.
Both vertex position and energy were calibrated by deploying a $^{252}$Cf source and using the 8~MeV peak in the energy spectrum resulting from thermal neutron capture on Gd (nGd).
The reconstruction performance can be found in~\cite{cite:PAC31} for nGd events and~\cite{cite:HJKPS2020} for events with up to 60~MeV using Michel electrons.

The selection criteria and estimated efficiency are given in Table~\ref{tab:selec_eff} and a more detailed description is found in~\cite{cite:TDR}. The selection criteria applied in this analysis are nearly identical to those in~\cite{cite:TDR}.

\begin{figure}
    %\centering
    \includegraphics[width=0.48\textwidth]{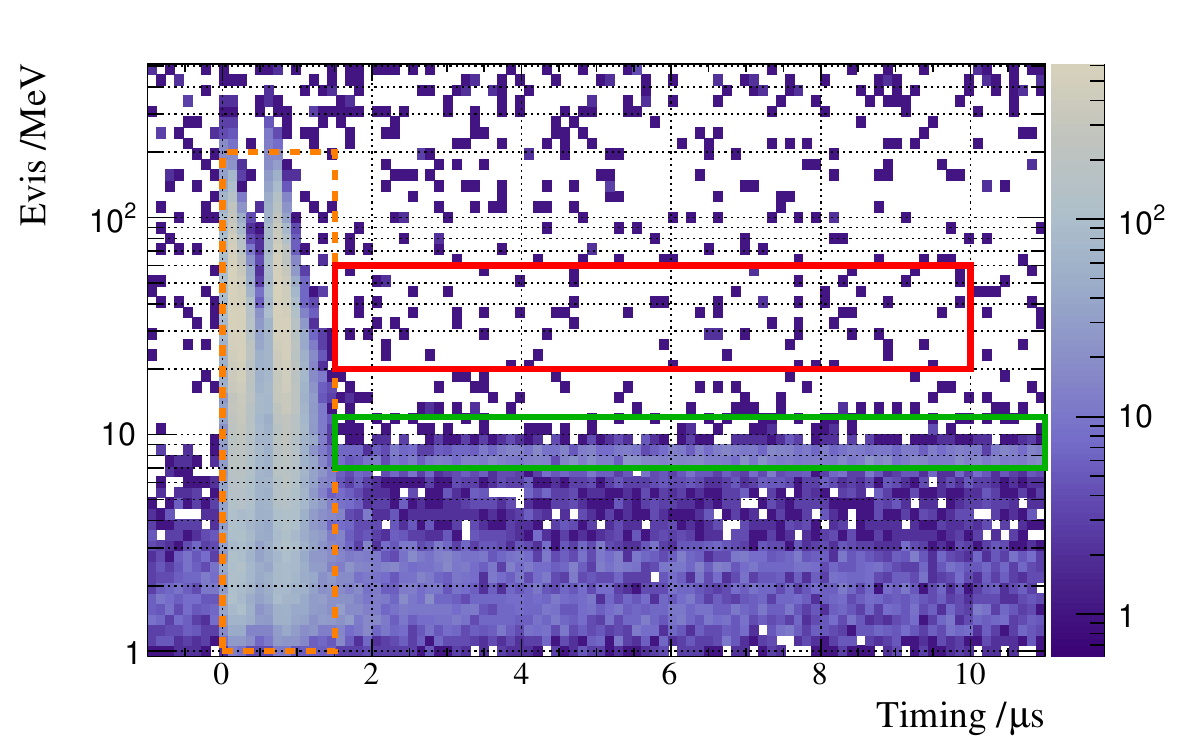}%
    \caption{A two-dimensional distribution of energy and timing before
    event selection, used to demonstrate the IBD selection region. The selected regions for the IBD prompt signal from positron (red box), the IBD delayed signal from gamma-rays resulted in thermal neutron capture on Gd (green box) and the beam on-bunch event (orange dashed box) are overlaid. 
    Note that the events are shown around the prompt signal timing region. There are two event clusters within 0 to 1.5 \micro, which reflect the proton beam structure of the MLF. They are caused by neutrons produced at the mercury target.
    One can see that the IBD prompt signal region is well separated from the on-bunch region.
    The events in the IBD delayed signal region must also satisfy $\Delta t_{\mathrm{p-d}} > 0$ in the actual delayed coincidence.
    \label{fig:e_t_def}}
\end{figure}

The prompt signal of IBD candidates is selected using a time difference from the beam collision timing ($\Delta t_{\mathrm{beam-p}}$) and its energy ($E_{\mathrm{p}}$).
We applied the following requirements; $1.5\le \Delta t_{\mathrm{beam-p}}\le 10$ \micro~and $20\le E_{\mathrm{p}}\le 60 $ MeV, in order to fully cover $\mu$DAR neutrinos from the mercury target.
The timing selection rejects beam-induced fast neutrons in the beam on-bunch timing ($0 \le \Delta t_{\mathrm{beam-p}}\le 1.5~\mathrm{\mu s}$) as well as neutrino backgrounds from kaon and pion decay whose lifetimes are 12~ns and 26~ns, respectively.
Fig.~\ref{fig:e_t_def} demonstrates the energy and timing selection windows in a two-dimensional distribution of energy and timing. The red box represents the prompt signal region and the orange dashed box displays the region of beam-induced fast neutron events.

In order to identify neutron captures on Gd, the 8 MeV peak of the delayed signal energy ($E_{\mathrm{d}}$) is selected using a requirement of $7\le E_{\mathrm{d}}\le 12 $ MeV~\cite{cite:SRBG}.
In addition, the conditions on the time and spatial difference between the prompt and delayed signal, $\Delta t_{\mathrm{p-d}} \le 100~\mathrm{\mu s}$ and $\Delta \mathrm{VTX}_{\mathrm{p-d}} \le 60$ cm, were used for correlated event selection. The range of digitized waveforms obtained using the kicker trigger limited the range of $\Delta t_{\mathrm{p-d}}$, and thus the selection was replaced with $\Delta t_{\mathrm{p-d}} \le 25~\mathrm{\mu s}$.

There are nGd events associated with fast neutrons induced by the beam contributing to the IBD delayed signal as an accidental background~\cite{cite:BG2014}.
Because these nGd events spatially correlate with an activity made by beam-induced fast neutrons, we can reject them the spatial correlation.
In particular,
we applied a requirement on the spatial difference between on-bunch event and delayed candidates, $\Delta \mathrm{VTX}_{\mathrm{OB-d}}\ge 110$ cm.
The distribution of $\Delta \mathrm{VTX}_{\mathrm{OB-d}}$ and the efficiency estimation can be found in~\cite{cite:SRBG}.
The on-bunch event tagging condition is set to $0 \le \Delta t_{\mathrm{beam-OB}} < 1.5~\mathrm{\mu s}$ and $1 \le E_{\mathrm{OB}}\le 200$ MeV.

Note that a nitrogen purge of the Gd-LS/LS was not performed. Thus, the PSD was unavailable in this commissioning run data.
The timing veto, a selection using a likelihood based on $\Delta t_{\mathrm{beam-p}}$ and $\Delta t_{\mathrm{p-d}}$~\cite{cite:SRBG}, was also not applied due to the limited waveform width.

\begin{table}[t]
    \centering
    \caption{The IBD selection criteria and their efficiencies in the \JSNS experiment~\cite{cite:TDR}. Note that due to the data taking strategy in the commissioning run we applied $\Delta t_{\mathrm{p-d}}\le 25$~\micro to the commissioning run data for background estimation. \label{tab:selec_eff}}
    \vspace{3pt}
    \begin{tabular}{cc}\hline
        Requirement & Efficiency /\% \\\hline
        $1.5\le \Delta t_{\mathrm{beam-p}}\le 10$~\micro    & 62 \\%\hline
        $20\le E_{\mathrm{p}}\le 60$~MeV                    & 92 \\%\hline
        $7\le E_{\mathrm{d}}\le 12$~MeV                     & 71 \\%\hline
        $\Delta t_{\mathrm{p-d}}\le 100$~\micro             & 93 \\%\hline
        $\Delta \mathrm{VTX}_{\mathrm{p-d}}\le 60$~cm       & 96 \\%\hline
        $\Delta \mathrm{VTX}_{\mathrm{OB-d}}\ge 110$~cm     & 98 \\%\hline
        PSD                                                 & 99 \\%\hline%\hline
        Timing veto                                         & 91 \\\hline
        Total                                               & 32 \\\hline
    \end{tabular}
\end{table}

\begin{figure}
    %\centering
    \includegraphics[width=0.5\textwidth]{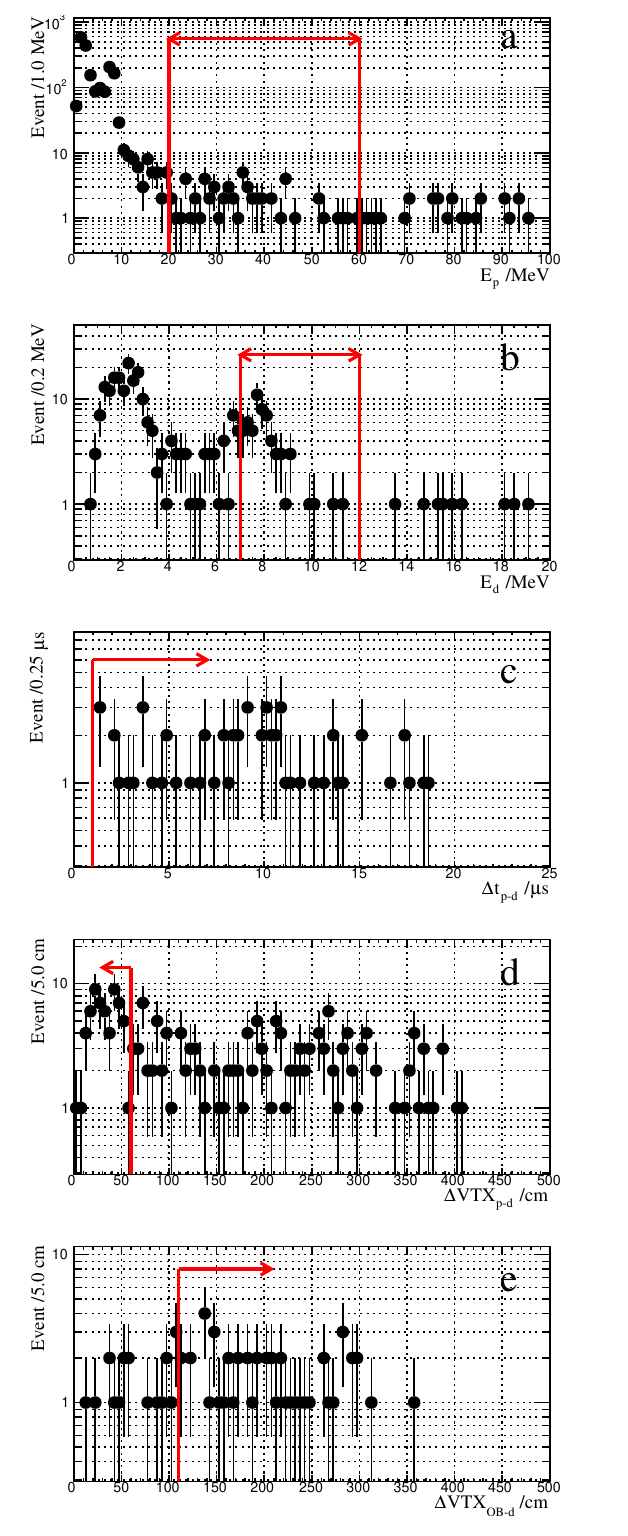}%
    \caption{The distributions of the variables used in the IBD selection: (a) $E_{\mathrm{p}}$, (b) $E_{\mathrm{d}}$, (c) $\Delta t_{\mathrm{p-d}}$, (d) $\Delta \mathrm{VTX}_{\mathrm{p-d}}$ and (e) $\Delta \mathrm{VTX}_{\mathrm{OB-d}}$. Each distribution has all of the selection criteria applied, except for the selection on the variable displayed. The red arrows show the selection criteria of each variable.
    \label{fig:result_ibd}}
\end{figure}

The IBD selection results in the distributions of $E_{\mathrm{p}}$ (a), $E_{\mathrm{d}}$ (b) and $\Delta \mathrm{VTX}_{\mathrm{p-d}}$ (c), $\Delta t_{\mathrm{p-d}}$ (d) and $\Delta \mathrm{VTX}_{\mathrm{OB-d}}$ (e) as shown in Fig.~\ref{fig:result_ibd}, respectively. The regions indicated by the red arrows represent the selected event candidates.
As a result, $59 \pm 8$ correlated event candidates remain within the IBD selection region for the sterile neutrino search.

Taking an efficiency difference between the beam-off data obtained by the self trigger and the beam-on data with a 25~\micro~FADC window into consideration, the expected number of cosmic-induced fast neutron events in the IBD selection is $55.9 \pm 4.3$.
This is consistent with the number of the correlated events obtained above.
The remaining background rates are mainly produced by cosmic-induced fast neutrons, with a measured rate of $(2.45 \pm 0.20) \times 10^{-5}$~/spill.
By assuming 99\% neutron rejection with the PSD capability, the cosmic-induced fast neutron event rate will be comparable with the expected event rate of the signal IBD at the best-fit oscillation parameter from the LSND experiment.
Therefore, it is crucial to construct a particle identification technique with the PSD capability for fast neutron rejection.

\section{Pulse shape discrimination}
\label{psd}

As described above, we have found a large number of background events caused by cosmic-induced fast neutrons for the sterile neutrino search.
Thus, diisopropylnaphthalene (DIN, $\mathrm{C}_{16}\mathrm{H}_{20}$) was dissolved into the Gd-LS in order to enhance the PSD capability.
DIN is commercially available and widely used as a base solvent of organic liquid scintillator. Several neutrino experiments using a liquid scintillator detector have adopted it and achieved good PSD capability~\cite{cite:NEOS, cite:PROSPECT}.
Approximately 8\% by volume of DIN was dissolved into the Gd-LS at the beginning of the first physics run from January 2021.
Nitrogen purging was performed before data taking which led to 10\% increase in light yield as a result.
The test data obtained using the self trigger in the first run was used to investigate the PSD capability.

We developed a PSD algorithm based on likelihood discrimination using probability density functions (PDFs) associated with the waveform.
The PDFs were constructed by accumulating the ratio of a peak count to a count at each sample of digitized waveforms for each PMT.
We defined a log-likelihood for PSD with a weight of the total charges representing photon statistics as follows:
\begin{equation}
    \log \mathcal{L}_{\alpha} = \sum^{96}_{i = 0} Q^{\mathrm{total}}_i \sum^{248}_{j = 40} \log( P^{\alpha}_i (Q^{\mathrm{ratio}}_{ij}) ) \quad (\alpha = \mathrm{sig, bkg}),
\end{equation}
where $Q^{\mathrm{total}}_i$ represents the observed charge on the $i$-th PMT, and $P^{\alpha}_i$ is the probability of the charge ratio $Q^{\mathrm{ratio}}_{ij}$ of the $j$-th sample of the digitized waveform of the $i$-th PMT. The first 39 samples were used for a baseline calculation and thus ignored in computing the likelihood value.
Two likelihood values representing signal-like ($\mathcal{L}_{\mathrm{sig}}$) and fast neutron background-like ($\mathcal{L}_{\mathrm{bkg}}$) were computed, and the log-likelihood difference of them, defined as a score, was then used for the event selection.

In parallel, another method for PSD was developed which utilizes a Convolutional Neural Network (CNN), a deep learning algorithm commonly used in the computer vision area.
The same data in the likelihood method are put in as a 96-channel image with $1 \times 208$ pixels (FADC sampling points).
A CNN structure composed of three blocks of convolution, max-pooling, ReLU, Batch Normalization and Dropout layers~\cite{cite:LeNet, cite:AI-Saffar} are used for the pattern recognition in the waveform image.
Two fully-connected layers composed by ReLU and Dropout layers are followed after the convolution block for the signal-to-background classification.
The model is trained to minimize the binary cross-entropy loss function, with the values closer to 1.0 corresponding to signals and 0.0 to the backgrounds.
The kernel size of the first convolution layer was chosen to be 11 for optimal performance of the signal-to-background classification.

In this study, we used the data samples of Michel electron events associated with cosmic muon and cosmic-induced fast neutrons for the signal-like and the background control sample, respectively, instead of Monte-Carlo simulation. The selected energy range is identical in terms of the IBD prompt signal definition (20 - 60 MeV).
As a starting point of the PSD study, the event vertices around the center of the detector ($r < 60$~cm and $|z| < 60$~cm) were selected.
We selected data files with odd ID numbers for the PDF construction and training the CNN model, and even ones for a performance evaluation.

\begin{figure}
    \centering
    \begin{minipage}{0.5\textwidth}
        \centering
        \includegraphics[width=0.98\textwidth]{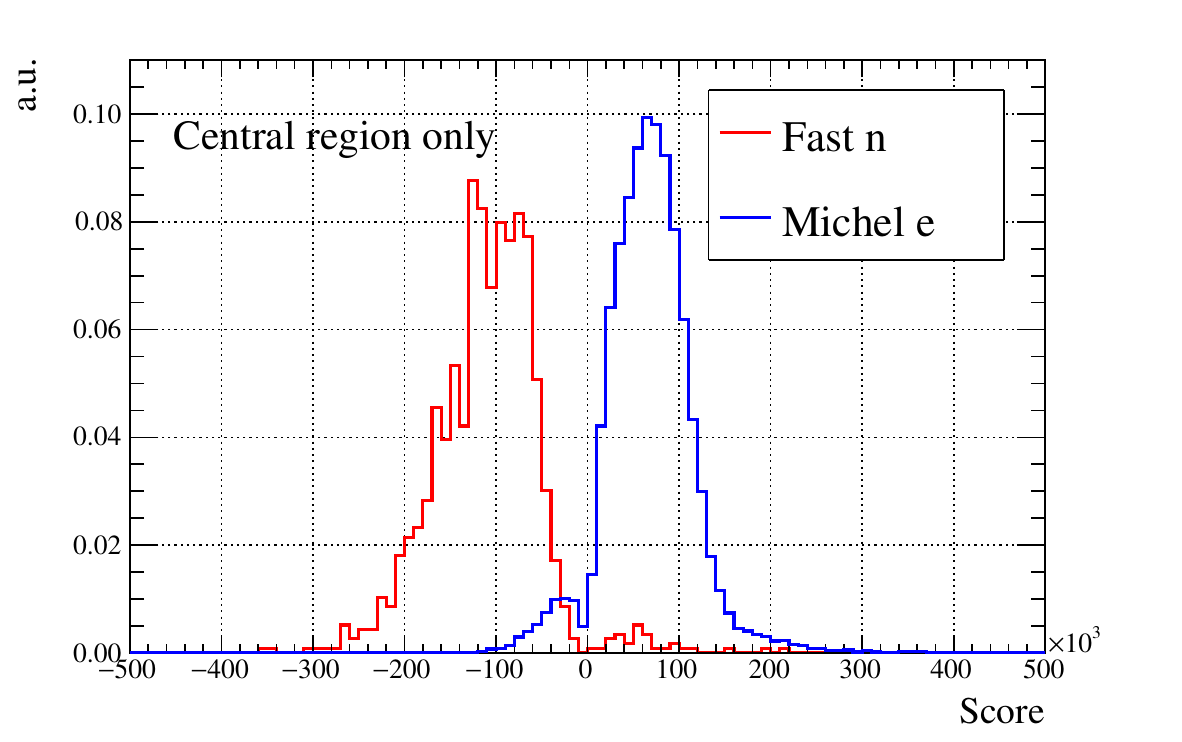}%
    \end{minipage}
    \centering
    \begin{minipage}{0.5\textwidth}
        %\centering
        \hspace{-4pt}
        \includegraphics[width=0.98\textwidth]{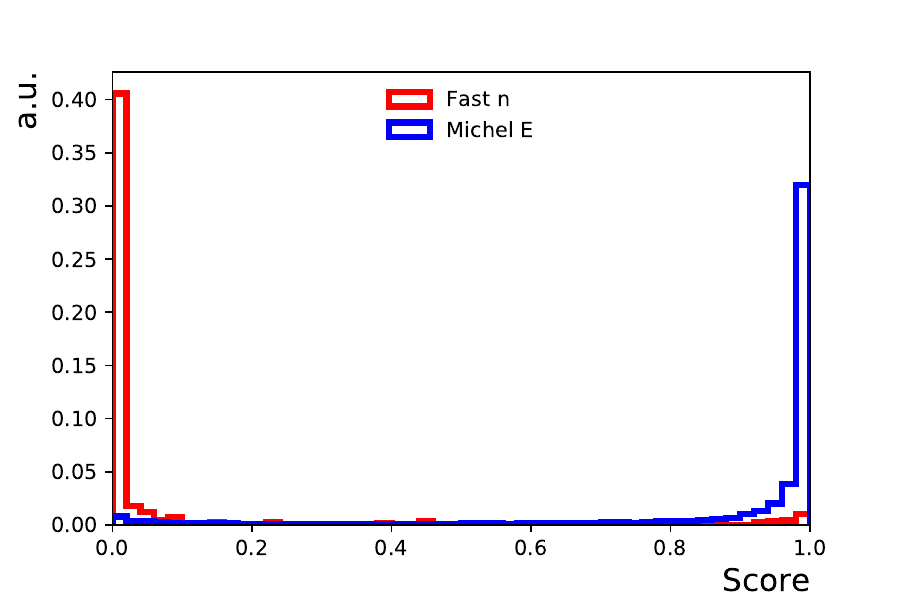}%
    \end{minipage}
    \caption{
        Top: The likelihood score distributions of the Michel electron sample (blue) and the fast neutron sample (red). The Michel electron sample represents the distribution of the neutrino signal.
        Bottom: The CNN score distributions of the Michel electron sample (blue) and the fast neutron sample (red).
        The label ``Central region only" indicates that the event vertices around the detector center ($r < 60$~cm and $|z| < 60$~cm) are used.
        \label{fig:psd}
    }
\end{figure}

Fig.~\ref{fig:psd} shows the result of the PSD performance of the likelihood (top) and the CNN method (bottom), respectively.
The likelihood method achieved fast neutron rejection of ($97.4 \pm 0.5$)\% in taking the likelihood score $\ge 0$ with a signal efficiency of ($94.2 \pm 2.6$)\%.
The performance of the CNN method was similar to the result of the likelihood method with a cut on the CNN score $\ge 0.17$.
Therefore, we have obtained close performance to the experimental requirement (99\% rejection), given an impurity of the fast neutron control sample coming from accidental coincidence caused by cosmic-induced gamma-ray leading to a signal-like pulse shape. 
An understanding of the impurity will improve the fast neutron rejection efficiency in maintaining the signal efficiency.

The \JSNS experiment performed data taking with the improvement against the dominant background for the sterile neutrino search from the physics run (January to June 2021), and obtained high statistics data equivalent to $1.45 \times 10^{22}$~POT.
According to the study on the PSD capability as a function of DIN concentration~\cite{cite:DIN}, further improvement on the PSD capability beyond the neutron rejection efficiency of 99\% can be anticipated by an increase in DIN concentration from 8 to 10\%, performed in June 2021~\cite{cite:detector}.

\section{Summary}
\label{summary}
\JSNS is a sterile neutrino search experiment aiming to directly test the positive result of the LSND experiment using a decay-at-rest neutrino source at the MLF and a Gd-LS detector. 

We performed the commissioning run of ten days in June 2020.
As a result of the IBD selection to the commissioning run data, we observed $59 \pm 8$ correlated event candidates in the signal region of the sterile neutrino search. 
This is consistent with the cosmic-induced fast neutron rate measured based on the beam-off data.

In order to reduce the cosmic-induced fast neutron background, 8\% in volume of DIN was dissolved into the Gd-LS to improve the PSD capability for first physics run starting January 2021.
We also developed two independent PSD methods based on a likelihood using waveform PDFs and machine learning with a CNN.
Using a control sample of Michel electrons and fast neutron events in the central region of the detector,
we achieved a fast neutron rejection of ($97.4 \pm 0.5$)\% with a signal efficiency of ($94.2 \pm 2.6$)\% if we require a likelihood score $\ge 0$.
The CNN method shows a similar performance to the likelihood method with a requirement of the CNN score $\ge 0.17$.
Further improvements on the PSD capability can be expected with an increase in DIN concentration as well as an understanding of the impurity in the control samples.
The \JSNS experiment will start data taking with the improvements from the coming physics run in 2021/2022.

\begin{acknowledgements}
\sloppy
We thank the J-PARC staff for their support. 
We acknowledge the support of the Ministry of Education, Culture, Sports, Science, and Technology (MEXT) 
and the JSPS grants-in-aid: 16H06344, 16H03967 and 20H05624, Japan.
This work is also supported by the National Research Foundation of Korea (NRF): 2016R1A5A1004684, 2017K1A3A7A09015973, 2017K1A3A7A09016426, 2019R1A2C3004955, 2016R1D1A3B02010606, 2017R1A2B4011200, 2018R1D1A1B07050425, 2020K1A3A7A09080133 and 2020K1A3A7A09080114.
Our work has also been supported by a fund from the BK21 of the NRF.
The University of Michigan gratefully acknowledges the support of the Heising-Simons Foundation. 
This work conducted at Brookhaven National Laboratory was supported by the U.S. Department of Energy under Contract DE-AC02- 98CH10886. 
The work of the University of Sussex is supported by the Royal Society grant no. IESnR3n170385.
We also thank the Daya Bay Collaboration for providing the Gd-LS, the RENO collaboration for providing the
LS and PMTs, CIEMAT for providing the splitters, Drexel University for providing the FEE circuits and Tokyo Inst. Tech for providing FADC boards.
\end{acknowledgements}



\end{document}